\documentclass[onecolumn,groupedaddress,nofootinbib,12pt]{revtex4-1}
\usepackage{amsmath}
\usepackage{graphicx}
\usepackage{sistyle}
\usepackage{gensymb}
\usepackage{bm}
\usepackage{natbib}
\usepackage{dsfont}
\usepackage{epsfig}
\usepackage{feynmf}
\usepackage{blindtext, rotating}
\usepackage{mathtools}
\usepackage{subcaption}
\usepackage{amsfonts}

\begin{document}
\title{Quantum Effects in Gravity Waves}
\author{Thiago Guerreiro}
\email{barbosa@puc-rio.br}

\affiliation{Department of Physics, Pontif\'icia Universidade Cat\'olica, Rio de Janeiro, Brazil}

\begin{abstract}
We discuss the quantum mechanical description of a gravitational wave interacting with a cavity electromagnetic field.
Quantum fluctuations of the gravitational vacuum induce squeezing in the optical field. Moreover, this squeezing experiences revivals, a purely quantum effect. Measuring these gravitationally induced revivals, although out of reach from experiments, would provide evidence on the quantum nature of gravity. We also discuss the quantum mechanical treatment of the interaction between coherent and squeezed gravitational wave states and a gravity wave detector. In the case of a coherent gravitational wave, we reproduce the result from the classical theory with a quantum mechanical calculation. The case of a squeezed gravity wave is not calculable within the classical theory, and could provide evidence on the quantum nature of gravity.
\end{abstract}


\maketitle

Gravitational wave astronomy is now a reality \cite{LIGO2019}. 
In parallel, important developments in quantum technology have enabled the precise control and measurement of complex quantum states \cite{Schnabel2019, Arute2019}. Despite these advances in our knowledge of gravity and quantum mechanics, combining both ideas in a quantum theory of the gravitational field remains a daunting task. In this letter, we explore whether gravitational wave astronomy and quantum technologies can be jointly exploited to shed light on the quantum mechanical nature of gravity.


The question of whether gravity has a description in terms of quantum fields is a long-standing one. Differently from electromagnetism \cite{Bohr1933}, the quantization of matter does not imply quantization of gravity \cite{Bronstein2012}. On top of that it is well known that quantum mechanics imposes limits to the measurement of spacetime geometry \cite{Osborne1949}. In particular, if an object is sufficiently microscopic, measuring the curvature of spacetime generated by the object might be impossible \cite{Wigner1967, Lloyd2012}. These and other arguments raise the question of whether gravity needs to be quantized in the first place \cite{Feynman2003, Diosi1987, Penrose1996}. Moreover, quantum corrections to gravity typically scale as powers of the Planck length, an exceedingly small number \cite{Donoghue1994} probably out of reach from conceivable experiments. Related to that is the question on the existence of gravitons, elementary excitations of the gravitational field \cite{Dyson2012}. Measuring gravitons is a challenging task which remains unachieved, despite some ideas in cosmology \cite{Wilczek2014}. 

One interesting possibility is to use quantum mechanics experiments to look for a witness of the `quantumness' of gravity \cite{Marletto2017, Carney2019, Blencowe2013, Pikowski2015, Belenchia2019}. The present letter goes along analysing that direction. Building on the Hamiltonian presented in \cite{Pang2018} for the interaction of a gravitational wave with a cavity quantized electromagnetic field we 
discuss a conceptually simple experiment which uses only coherent states and requires no superpositions of energy configurations in different locations that could in principle evidence quantum mechanical effects deriving from the interaction of matter with the gravitational field and gravitational waves. 
\begin{figure}[ht!]
\includegraphics[scale=0.8]{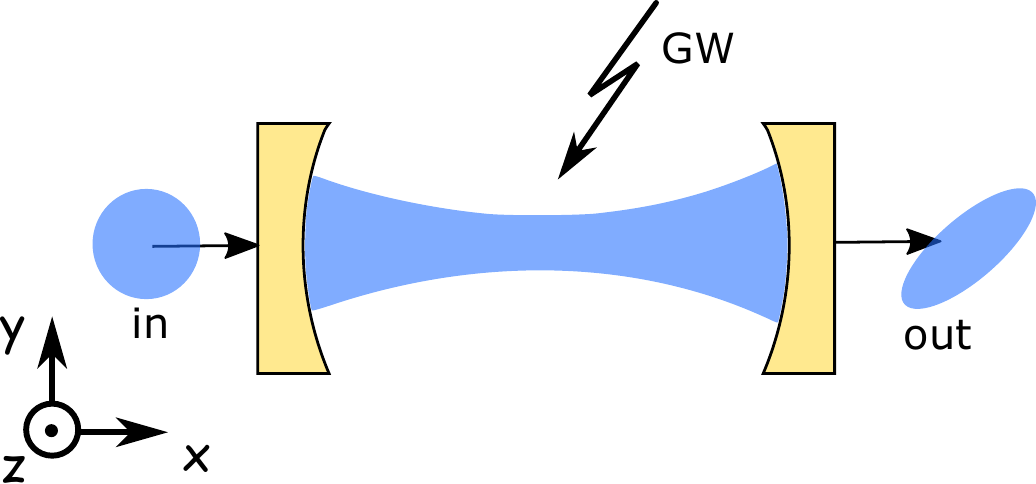}
\caption{\label{Interferometer1} Proposed experiment.
}
\end{figure}
After thoroughly discussing the Hamiltonian we perform a quantum mechanical calculation of the interaction between a cavity electromagnetic field and the gravitational vacuum. We show that quantum gravity induces oscillations in the electromagnetic field quadrature and variance. In other words, gravity induces (anti-)squeezing in the field and in addition the squeezing exhibits revivals. Although electromagnetic squeezing in itself can be produced via the interaction of light with a classical oscillator, and hence is not sufficient to evidence quantum behavior, \textit{revivals} of squeezing are known to be a purely quantum effect and hence provide a witness of non-classicality \cite{Ma2018}. Hence, it is in principle possible to obtain evidence on the quantum nature of gravity through quantum optics, despite these are out of reach from conceivable experiments. 

We then consider the interaction of a cavity electromagnetic field with coherent and squeezed gravitational wave states. We perform quantum mechanical calculations of the time dependency of the optical field quadratures. In the coherent gravitational wave case, we recover the result from the classical theory, namely that a passing gravitational wave induces phase oscillations in the light field as observed by LIGO. 
The interaction of a squeezed gravitational wave with the cavity field can only be calculated within quantum theory, has no classical analogue and to the best of our knowledge, this is the first time such calculation has been reported. Squeezed gravitational waves might be interesting in cosmology and could provide a route towards experimental evidence on the quantum nature of gravity. 

The Universe is embedded in a thermal gravitational wave background \cite{Allen1996}. For that reason we also consider the effect of such a thermal background to the single mode calculation. Taking the temperature of these thermal gravitons to be the expected value for the cosmic gravitational wave background we can show that corrections to the pure state calculation are negligible. 
%



\section{Hamiltonian}

Consider a single-mode Fabry-P\'erot cavity with massive mirrors. It has been suggested that a cavity of this kind with an optically levitated nanosphere inside can serve as an optomechanical gravitational wave detector \cite{Arvanitaki2013}. The passing of a gravitational wave alters the position of the nanosphere inside the cavity, which induces a change in the cavity's resonance frequency. It is expected that a similar effect happens with a cavity in which one of the mirrors is free to move as the wave passes through. Consider such a cavity with length $ \ell_{0} $ in which one of the mirrors is weakly bound by a harmonic potential. As seen from the locally Lorentz reference frame of one of the mirrors, the passing of a weak gravitational wave will change the cavity length to $ \ell_{0} (1 + \frac{1}{2} h) $. Following \cite{Arvanitaki2013} we note that this induces an optomechanical-like coupling between the gravitational wave and the moving mirror. The resonance frequency of the cavity fundamental mode will then be
\begin{eqnarray}
\omega = \dfrac{n\pi}{\ell_{0} (1 + \frac{1}{2} h)} = \omega_{0} \left( 1 - \dfrac{1}{2} h + \mathcal{O}(h^{2}) \right) 
\end{eqnarray}
where $ \omega_{0} = n\pi / \ell_{0} $ and $ n $ is an integer.
It is then expected that this leads to an effective coupling between a gravitational wave strain $ h $ and the cavity mode through the electromagnetic Hamiltonian $ \omega a^{\dagger} a $. Natural units are used throughout. 

The Hamiltonian describing the interaction of a test mass (a mirror in a weak harmonic potential), the cavity electromagnetic field and a gravitational wave has been analysed in \cite{Pang2018}. The total Hamiltonian consists of the terms,
\begin{eqnarray}
\hat{H} = H^{(0)}_{q} + H^{(0)}_{EM} +  H^{(0)}_{GW} + H_{OM} + H^{\rm int}_{GW}
\end{eqnarray}
where $ H^{(0)}_{q}  $, $ H^{(0)}_{EM} $ and $ H^{(0)}_{GW} $ are the free Hamiltonians for the test mass, the electromagnetic field and the gravitational field, respectively, $ H_{OM} $ is an optomechanical coupling between the mirror and the electromagnetic field and $ H^{\rm int}_{GW} $ the interaction between the electromagnetic field and the gravitational wave. The optomechanical term $ H_{OM} $, sometimes also referred to as the radiation pressure term is given by \cite{Law1995},
\begin{eqnarray}
H_{OM} = - \dfrac{\omega_{0}}{\ell_{0}} \left(   \dfrac{1}{2M \omega_{m}}  \right) ^{1/2} a^{\dagger} a  (c + c^{\dagger})
\end{eqnarray}
where $ \ell_{0} $ is the cavity length, $ \omega_{m} $ the mirror's resonance frequency, $ M $ the mirror's mass and $ (a, a^{\dagger}) $ and $ (c, c^{\dagger}) $ the electromagnetic and mirror's annihilation and creation operators, respectively. Note this term can be made very small by choosing very large mass mirrors. 


Finally, the term $ H^{\rm int}_{GW} $ corresponds to the interaction between a canonically quantized metric perturbation $ \hat{h}_{ij}(x,t) $ in the TT (transverse traceless) gauge and the cavity field. The perturbation assumes the form  \cite{Oniga2016}
\begin{eqnarray}
\hat{h}_{ij}(x,t) = \sum_{\lambda} \int \dfrac{d^{3} k}{\sqrt{(2\pi)^{3}}} \left(   \sqrt{\dfrac{8\pi G}{k}} \mathfrak{b}_{k}^{\lambda} \ e_{ij}^{\lambda}(k) \ e^{i(kx - \omega t)} + \mathrm{h.c.}  \right)
\end{eqnarray}
where $ \omega = k $, $ \lambda = +, \times $ are the wave polarizations, $  \mathfrak{b}_{k}^{\lambda} $ ($ \mathfrak{b}_{k}^{\lambda \dagger} $) the canonical annihilation (creation) operator for polarization mode $ \lambda $ satisfying
\begin{eqnarray}
\left[  \mathfrak{b}_{k}^{\lambda} , \mathfrak{b}_{k'}^{\lambda ' \dagger}   \right] = \delta_{\lambda \lambda '} \delta^{(3)}(k,k') \ , \ \ \left[  \mathfrak{b}_{k}^{\lambda \dagger} , \mathfrak{b}_{k'}^{\lambda ' \dagger}   \right] = \left[  \mathfrak{b}_{k}^{\lambda} , \mathfrak{b}_{k'}^{\lambda ' }   \right] = 0
\end{eqnarray}
and $ e_{ij}^{\lambda}(k) $ the unit TT tensors for mode $k $. 

Starting from the Einstein-Hilbert action coupled to the Maxwell stress energy tensor, interaction of the metric perturbation $ \hat{h}_{ij}(x,t) $ with a cavity electromagnetic field can be derived. For a $ + $ polarized gravitational wave propagating along the $ \hat{z} $ direction, \textit{perpendicularly} to the cavity axis ($ \hat{x}$-direction shown in figure~\ref{Interferometer1}) \cite{Bassi2017}, the Hamiltonian describing coupling of the wave to the electromagnetic field takes the form
\begin{eqnarray}
H^{\rm int}_{GW} = - \dfrac{\omega_{0}}{4} a^{\dagger} a \int \dfrac{d^{3} k}{\sqrt{(2\pi)^{3}}} \left(   \sqrt{\dfrac{8\pi G}{k}} \mathfrak{b}_{k} + \mathrm{h.c.}  \right)
\label{H_int_canonical}
\end{eqnarray} 
where we have assumed that the gravitational wavevector component along the cavity axis $ k_{x} $ satisfies  $ k_{x}\ell_{0} \ll 1 $. Details of the Hamiltonian derivation can be found in \cite{Pang2018, Oniga2016, Bose2017}.
Terms involving the interaction of the helicity of gravity with that of photons have been neglected, as the effect of these terms happens on a time scale much shorter that the effects we are interested in \cite{Pang2018}. These effects can also be averaged by using circularly polarized light.

The interaction Hamiltonian \eqref{H_int_canonical} describes a gravitational optomechanical coupling. In the following, we will consider some effects of this Hamiltonian on the electromagnetic field in the presence of different quantum states of the gravitational field.
To simplify we will neglect the free evolution of the test mass $ H^{(0)}_{q} $ and the radiation pressure term $ H_{OM} $; this is justified in the limit that the cavity mirror has a very large mass, and these terms effectively decouple from the electromagnetic field dynamics. 

For simplicity we wish to express the operator \eqref{H_int_canonical}  in a discrete version by introducing a quantization volume $ V $ to avoid infrared divergences. This is a common device in the quantization of the electromagnetic field in quantum optics (see chapter 1, eq. (1.1.23) of \cite{Scully1997} for example). At the end of the desired calculations, the quantization volume can be taken to be infinite and the discrete set of modes can be taken to a continuum limit.   
Note that in \eqref{H_int_canonical}, the quantity $ \sqrt{8\pi G / k} $ has units of (length)$^{3/2}$, $ \left[ \sqrt{8\pi G / k} \right] = L^{3/2} $. Also, $ \left[ d^{3} k \right] = L^{-3} $. This implies that the annihilation (creation) operator $ \mathfrak{b}_{k} $ ($ \mathfrak{b}_{k}^{\dagger} $) must have dimension $ \left[ \mathfrak{b}_{k} \right] = L^{3/2} $ of a density in momentum space. We introduce the dimensionless annihilation operator $ \mathfrak{b}_{k} = b_{k} / (1/ \sqrt{V}) = \sqrt{V}  b_{k} $, where $  b_{k} $ is now a dimensionless quantity. In the discrete limit, we have  $ d^{3} k \rightarrow 1 / V $ and $ (2\pi)^{-3/2}\int \rightarrow \sum $. This leads to the discrete version of \eqref{H_int_canonical}, or Riemann sum form of the interaction Hamiltonian,
\begin{eqnarray}
H^{\rm int}_{GW} = - \dfrac{\omega_{0}}{4} a^{\dagger} a \sum_{k} \left(   \sqrt{\dfrac{8\pi G}{kV}} b_{k} + \mathrm{h.c.}  \right)
\end{eqnarray}
Note that this interaction Hamiltonian has the same form as the one reported in \cite{Bose2017}. For completeness we take the discrete limit of the free gravitational Hamiltonian,
\begin{eqnarray}
H_{GW}^{(0)} = \int \dfrac{d^{3} k}{\sqrt{(2\pi)^{3}}} \Omega_{k} \  \mathfrak{b}_{k}^{\dagger} \mathfrak{b}_{k} \rightarrow \sum_{k}  \Omega_{k} \  b_{k}^{\dagger} b_{k}
\end{eqnarray}
where $ \Omega_{k} = \vert k \vert $ is the gravitational wave frequency in mode $ k $.

The effective interaction between the electromagnetic and gravitational fields is then
\begin{eqnarray}
\hat{H} = \omega_{0} a^{\dagger}a &+&\sum_{k} \Omega_{k} b^{\dagger}_{k} b_{k} \nonumber \\ 
&-& \dfrac{\omega_{0}}{4} a^{\dagger}a \sum_{k} \left(   \sqrt{\dfrac{8\pi G}{kV}} b_{k} + \mathrm{h.c.}  \right)
\end{eqnarray}
Here, $a^{\dagger}a $ counts the number of excitations in the electromagnetic field coming from the free electromagnetic Hamiltonian; similarly for $ b^{\dagger}_{k} b_{k} $ which counts the number of excitations in the gravitational field.

The quantity $ f_{k} =  \sqrt{\dfrac{8\pi G}{kV}}  $ can be interpreted as the single graviton strain for mode $ k $.
We define the normalized coupling strength $ g_{k} = \omega_{0} f_{k} / 4 $ and the dimensionless quantities $ q_{k} = g_{k} / \Omega_{k}  $. Note that
\begin{eqnarray}
 \mathrm{max} \lbrace q_{k} \rbrace = \omega_{0} \sqrt{\dfrac{8\pi G}{k_{\rm min}^{3} V}} \simeq \dfrac{\omega_{0}}{E_{\rm pl}} \label{max}
\end{eqnarray}


With the above definitions the Hamiltonian becomes
\begin{eqnarray}
\hat{H} = \omega_{0} a^{\dagger}a +\sum_{k}  \Omega_{k} b^{\dagger}_{k} b_{k} - \sum_{k} g_{k} a^{\dagger} a \ (b_{k}  + b^{\dagger}_{k})
\label{hamilton}
\end{eqnarray}
The effective coupling between the electromagnetic and gravitational field is proportional to the number of photons in the cavity. Note that each term in \eqref{hamilton} describes the standard interaction of cavity quantum optomechanics \cite{Asplemeyer2014} with the association \textit{oscillator position} $  \sim  $ \textit{one mode of the gravitational field}. 


We can now calculate the unitary evolution operator associated to \eqref{hamilton}. Because of the commutation relations between different gravitational modes,
\begin{eqnarray}
\left[   b_{k},   b_{k'}  \right]  = 0  \ , \  k\neq k'
\end{eqnarray}
the evolution operator is of the form
\begin{eqnarray}
U(t) = e^{-i\hat{H}t } = \prod_{k} U_{k} (t) 
\label{unitary}
\end{eqnarray}
where each unitary $ U_{k} (t)  $ acts jointly \textit{only} on the cavity mode and the gravitational wave mode $ k $, having the form \cite{Ma2018, Bose1997} (see the Appendix for a derivation)
\begin{eqnarray}
U_{k} (t)  = e^{iA_{k}(t) (a^{\dagger}a)^{2}} e^{q_{k} a^{\dagger} a  ( \gamma_{k}^{*} b_{k} - \gamma_{k} b^{\dagger}_{k}   )}
\label{Uk}
\end{eqnarray}
where $ \gamma_{k} = (1 - e^{i\Omega_{k} t}) $ and $ A_{k}(t) = 2 q_{k}^{2} \left(   \Omega_{k} t - \sin \Omega_{k} t  \right)   $. 
The Heisenberg equation for the oscillator's annihilation operator reads \cite{Ma2018}
\begin{eqnarray}
a(t) &=& \left(    \prod_{k} U_{k} (t)   \right)^{\dagger}  a    \left(    \prod_{k} U_{k} (t)   \right) = 
\\
&=& \prod_{k}  e^{i A_{k}(t)a^{\dagger}a} e^{iA_{k}(t)/2} e^{q_{k}  ( \gamma_{k}^{*} b_{k} - \gamma_{k} b^{\dagger}_{k}   )} a = 
\\
&=&  \prod_{k}  e^{i A_{k}(t)a^{\dagger}a} e^{iA_{k}(t)/2}  \mathcal{D}(-q_{k} \gamma_{k} ) \ a \label{a_final}
\end{eqnarray}
where $ \mathcal{D}(-q_{k} \gamma_{k} ) $ is a displacement operator acting on the gravitational wave mode $ k $.

\section{Vacuum state}

We now consider the gravitational field to be in the vacuum state. We are interested in calculating the field quadrature mean value
\begin{eqnarray}
\langle \tilde{\mathcal{E}}_{+} \rangle \equiv \dfrac{\langle \mathcal{E}_{+} \rangle}{\mathcal{E}_{s}} =  \langle a^{\dagger} \rangle +  \langle a \rangle \label{quadrature}
\end{eqnarray}
where $ \mathcal{E}_{s} = \sqrt{\omega_{0} / 2\ell_{0}^{3}} $, and $ \ell_{0}^{3} $ is the cavity mode-volume \cite{Davidovich1996}. We are also interested in the quadrature variance $ \Delta \tilde{\mathcal{E}}_{+} $, defined
\begin{eqnarray}
\Delta \tilde{\mathcal{E}}_{+}^{2}  = \langle \tilde{\mathcal{E}}_{+}^{2} \rangle - \langle \tilde{\mathcal{E}}_{+} \rangle^{2} \label{variance}
\end{eqnarray}
As well known in quantum optomechanics, each unitary operator \eqref{Uk} gives rise to squeezing in the cavity field, due to the quadratic ``Kerr-like'' factor $ (a^{\dagger}a)^{2} $. The displacement factor, $ e^{q_{k} a^{\dagger} a  ( \gamma_{k}^{*} b_{k} - \gamma_{k} b^{\dagger}_{k}   )} $, can be thought of as a linear optics term which changes the phase of the field but does not generate quadrature squeezing. 

To estimate \eqref{quadrature} and \eqref{variance} we consider the Heisenberg equation \eqref{a_final}, summed over all modes of the gravitational field,
\begin{eqnarray}
a(t) =  e^{i F(t)a^{\dagger}a} e^{iF(t)/2}   D(t) \ a
\end{eqnarray}
where
\begin{eqnarray}
F(t) &=& \sum_{k} A_{k}(t) 
\end{eqnarray}
and
\begin{eqnarray}
D(t) &=& \prod_{k} \langle       0_{k} \vert   \mathcal{D}( -q_{k} \gamma_{k}) \vert 0_{k}  \rangle =\exp \left( -\dfrac{1}{2} \sum_{k} q_{k}^{2} \vert  \gamma_{k} \vert^{2}  \right) \nonumber 
\\
\end{eqnarray}
Strictly speaking, $ F $ and $ D $ diverge. To obtain a finite result we revert to the continuum limit and introduce an infrared and an ultraviolet cut-off\footnote{We note that the introduction of an ultraviolet cut-off is consistent with condition $k_{x} l_{0} \ll 1$, used in the derivation of eq.(6). That condition signifies that the relevant modes that couple to a cavity with axis along the $ \hat{x} $ direction according to Hamiltonian (6) are plane waves propagating perpendicular to $ \hat{x} $. The ultraviolet cut-off is introduced for the remaining $ k_y, k_z $ components constituting $ k $, manifest as a cut-off in the modulus of $ k $.}, at the Hubble and Planck energies, respectively,
\begin{eqnarray}
F(t) &=& \int \dfrac{d^{3} k}{\sqrt{(2\pi)^{3}}} 2\omega_{0}^{2}  \left( \dfrac{ 8\pi G}{k^{3}} \right) \left( \Omega_{k} t - \sin \Omega_{k} t    \right)  \\
&\simeq & \left( \dfrac{\omega_{0}}{E_{\rm pl}}  \right)^{2} \left(  \int^{E_{\rm pl}}_{E_{IR}} dk  \right) t = \left( \dfrac{\omega_{0}}{E_{\rm pl}}  \right) \omega_{0} t
\end{eqnarray}
where we have taken the limit of large $ t $ (that is, consider $ t \gg E_{IR}^{-1} $ and neglect the bounded term $ \sin \Omega_{k} t  $  under the integral sign) and neglected pre-factors of order one.

The factor $ D $ acts as a damping factor, reducing the squeezing and its oscillations. For that reason we approximate the integral in the exponential by a worst-case value. The estimate for $ D $ reads
\begin{eqnarray}
D(t) &=& \exp \left( -\dfrac{1}{2} \int \dfrac{d^{3} k}{\sqrt{(2\pi)^{3}}} \omega_{0}^{2}  \left( \dfrac{ 8\pi G}{k^{3}} \right) \vert  \gamma_{k} \vert^{2} \right) \\
&\simeq & \exp \left( - \left( \dfrac{\omega_{0}}{E_{\rm pl}}  \right)^{2} \int^{E_{\rm pl}}_{E_{IR}} \dfrac{dk}{k} \right)  \\
&=& \exp \left( - \left( \dfrac{\omega_{0}}{E_{\rm pl}}  \right)^{2} \ln \left( \dfrac{E_{\rm pl}}{E_{IR}} \right)  \right) \equiv D
\end{eqnarray}
where we have taken the worst-case approximation by considering the bounded term $  \vert  \gamma_{k} \vert^{2} = 2(1 - \cos \Omega_{k}t) \sim 2 $ and neglected numerical factors of order one. Plugging in numbers, $ \ln (E_{\rm pl} / E_{IR}) = \ln( \ell_{U} / \ell_{\rm pl} ) \simeq \ln 10^{62} \simeq 62 $.

For the cavity field in a coherent state $ \vert \alpha \rangle $, with $ a \vert \alpha \rangle = \alpha \vert \alpha \rangle $, assuming for simplicity $ \alpha $ real, the quadrature is
\begin{eqnarray}
\langle \tilde{\mathcal{E}}_{+} \rangle &=& \alpha D \left( e^{i F(t) / 2} \langle \alpha \vert \alpha e^{iF(t)} \rangle + c.c. \right) \\
&=& 2 \alpha D e^{-\alpha^{2} \left( 1 - \cos F(t) \right) } \cos \left(   \dfrac{F(t)}{2} + \alpha^{2} \sin F(t)   \right) \nonumber
\end{eqnarray}

\begin{figure}[ht!]
\includegraphics[scale=0.6]{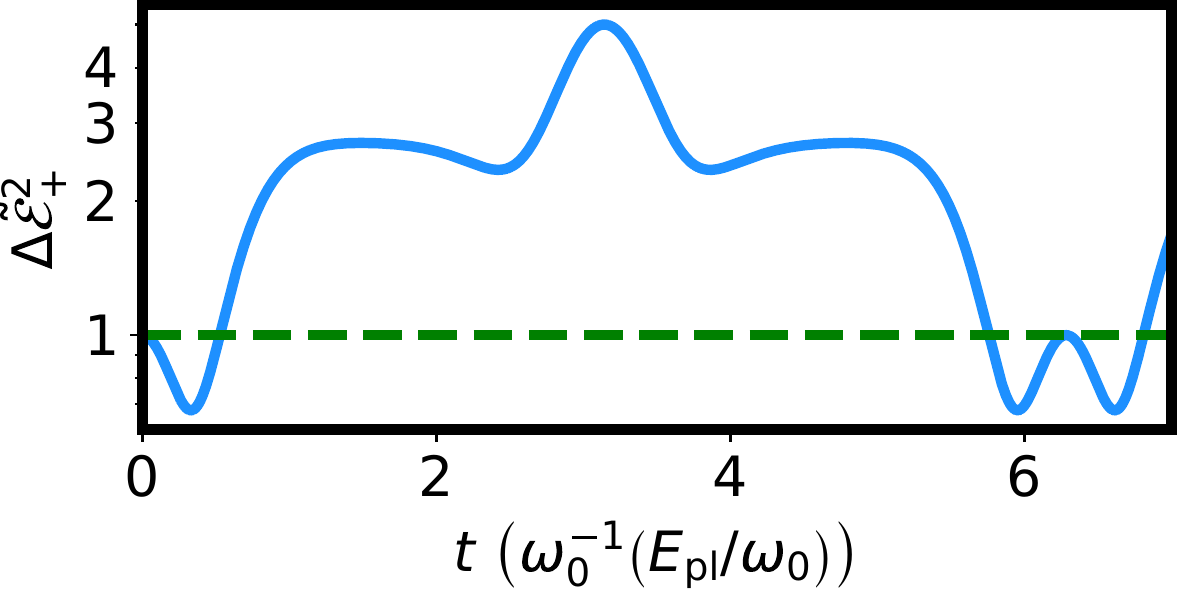}
\caption{\label{var} Numerical plot of variance as a function of time. For this plot the coherent state was taken to be $ \alpha = 1 $. For specific times, the variance falls below one, indicating the gravitational induced squeezing.
}
\end{figure}

The mean squared value of the quadrature is given by
\begin{eqnarray}
\langle \tilde{\mathcal{E}}_{+}^{2} \rangle &=& 2\alpha^{2} D^{2}  e^{-\alpha^{2} \left( 1 - \cos 2F(t) \right) } \cos \left(  F(t) + \alpha^{2} \sin 2 F(t)   \right) \nonumber
\\
&+& 2\alpha^{2} + 1
\end{eqnarray}
Combining the above results we obtain the variance
\begin{eqnarray}
&\ &\Delta \tilde{\mathcal{E}}_{+}^{2}  = \langle \tilde{\mathcal{E}}_{+}^{2} \rangle - \langle \tilde{\mathcal{E}}_{+} \rangle^{2} = \nonumber
\\
&\ & 2 \alpha^{2} D^{2}  e^{-\alpha^{2} \left( 1 - \cos 2F(t) \right) } \cos \left(  F(t) + \alpha^{2} \sin 2 F(t)   \right)  \nonumber
\\
&\ & - 2  \alpha^{2} D^{2}  e^{-2 \alpha^{2} \left( 1 - \cos F(t) \right) } \cos \left(  F(t) + 2 \alpha^{2} \sin  F(t)   \right) \nonumber
\\
&\ & 2\alpha^{2} \left(    1 -  D^{2}  e^{-2 \alpha^{2} \left( 1 - \cos F(t) \right) }   \right) + 1 \label{vacuum_var}
\end{eqnarray}
with the estimates for $ D $ and $ F(t) $ given above. 
For specific intervals of time, this variance falls below one \cite{Ma2018} as can be seen from the numerical plot in figure~\ref{var}. This implies that the optical field exhibits squeezing \cite{Lvovsky2014}. Numerically, the first minimum of $ \Delta \tilde{\mathcal{E}}_{+}^{2} $ occurs when $ (\omega_{0} / E_{\rm pl})^{2} \omega_{0} t_{0} \approx 0.33 $ at which point $ \Delta \tilde{\mathcal{E}}_{+}^{2} \approx 0.68 $. For optical frequencies the time of minimal variance corresponds to an astronomically long time of $ t_{0} \approx \SI{10^{41}}{s} $, out of reach from experiments. For higher energies of the field minimal squeezing occurs before, but only reaches a reasonable time as $ \omega_{0} $ approaches $ E_{\rm pl} $. 

From figure \ref{var} we also see the squeezing experiences revivals, which are known to be a purely quantum mechanical effect. These revivals would provide a witness of the quantum mechanical nature of the gravitational field, but once again, out of reach from experiments. 

For completeness, we consider the effect of a thermal bath of gravitons over the above pure state calculation. Following \cite{Ma2018}, note that a thermal background multiplies a damping factor to \eqref{vacuum_var} which is at most $  e^{-4q_{\rm max}^{2}(2\bar{n} + 1)} $ 
where $  \bar{n} = 1 / (e^{\hbar \Omega_{k}/k_{B}T} - 1)  $. For a gravitational wave mode of \SI{10}{Hz} (the expected peak of the cosmic spectrum) at the predicted temperature of the primordial gravitational wave background of $ T \sim \SI{1}{K} $ \cite{Allen1996}, this damping factor is approximately $ \exp(-10^{-47})\simeq 1 $. This is essentially saying that the decoherence due to the stochastic gravitational wave background is a negligible effect.
As a final remark regarding decoherence, we point out that any gravitational decoherence mechanism would be present in all quantum experiments, regardless of the quantum property one wishes to measure. If gravitational decoherence would impose a restriction to the observation of quantum properties of matter, that restriction would be present in any quantum experiment. 

\section{Coherent states}

We can use equations \eqref{Uk} and \eqref{a_final} to calculate the interaction of the cavity optical field with the gravitational field in any quantum state. We now consider a coherent gravitational wave, that is, one mode of the gravitational field will be populated by a coherent state, while all the others are taken to be in the vacuum state. It is safe to consider the gravitational wave to be in a pure state since dispersion and decoherence are negligible over distances up to the radius of curvature of the Universe \cite{MTW}. We also consider, for simplicity, monochromatic gravitational waves, which strictly speaking is not the case observed by gravitational wave detectors. Gravitational waves emitted from binary systems have a chirp in the frequency. Over short times, however, one may consider the waves as essentially monochromatic. Moreover, since we are working under the approximation of weak linearized gravity the chirp in the gravitational wave translates linearly to a first approximation to any imprinted fluctuations upon the electromagnetic field.

As we asserted in the previous section that the effects of squeezing and of the gravitational vacuum are very small within the reasonable time of an experiment, we will neglect all factors regarding vacuum expectation values in the quadrature of the field as very close to one. This amounts to approximating equation \eqref{Uk} by the linear optical term
\begin{eqnarray}
U_{k} (t) \simeq  e^{q_{k} a^{\dagger} a  ( \gamma_{k}^{*} b_{k} - \gamma_{k} b^{\dagger}_{k}   )}
\end{eqnarray}
and the Heisenberg equation by
\begin{eqnarray}
a(t) \simeq  \prod_{k}  \mathcal{D}(-q_{k} \gamma_{k} ) \ a
\label{a_final2}
\end{eqnarray}
Once again, we take the optical field to be in a coherent state $ \vert \alpha \rangle $, $ \alpha $ real, a single mode monochromatic gravitational wave $ k_{GW} $ in the coherent state represented by $ \vert \lambda e^{i\Omega_{GW} t} \rangle $ and all the remaining gravitational wave modes in the vacuum state $ \vert 0_{k} \rangle $ ($ k\neq k_{GW} $). 

Since \eqref{a_final2} is a tensor product of operators acting on independent modes and we are considering only product states, we can write the mean value of $ a $ as a product,
\begin{eqnarray}
\langle a(t) \rangle = \langle \alpha \vert a \vert \alpha \rangle \langle \Psi \vert \mathcal{D}(-q_{k_{GW}} \gamma_{\rm{gw}} ) \vert \Psi \rangle \prod_{k\neq k_{GW}} \langle 0_{k} \vert  \mathcal{D}(-q_{k} \gamma_{k} ) \vert 0_{k} \rangle \nonumber \\
\ \label{a2}
\end{eqnarray}
As before, each factor $ \langle 0_{k} \vert  \mathcal{D}(-q_{k} \gamma_{k} ) \vert 0_{k} \rangle $ is the overlap of a coherent state $ \vert -q_{k} \gamma_{k}  \rangle  $ with the vacuum state $ \vert 0_{k} \rangle  $, which can be written as,
\begin{eqnarray}
\langle       0_{k}    \vert -q \gamma_{k}  \rangle = e^{-q_{k}^{2} \vert \gamma_{k} \vert^{2} / 2}
\end{eqnarray}
The argument in the exponential is of order $ q_{k}^{2} $ and thus the above expression can be taken as very close to one with corrections starting at $ \mathcal{O}( \omega_{0}^{2} / E_{\rm pl}^{2}) $, as we showed in equation \eqref{max}. The expectation value of $ a(t)$ is
\begin{eqnarray}
\langle a(t) \rangle = \langle \alpha \vert a \vert \alpha \rangle \langle \Psi \vert \mathcal{D}(-q_{k_{GW}} \gamma ) \vert \Psi \rangle \label{Final_heisenberg}
\end{eqnarray}
where we redefine $ \gamma_{GW} \equiv \gamma = (1 - e^{i\Omega_{GW} t}) $ and $ q_{k_{GW}} \equiv q $. 

The mean quadrature is
\begin{eqnarray}
\langle \tilde{\mathcal{E}}_{+} \rangle &=& \alpha \left(   \langle \lambda e^{i\Omega_{GW}t} \vert \mathcal{D}(-q\gamma) \vert \lambda e^{i\Omega_{GW}t} \rangle  + c.c. \right) \\
&=& \alpha \left[     e^{\left(  -\frac{1}{2}q^{2} \vert \gamma \vert^{2} - i (\lambda q) \sin \Omega_{GW} t     \right)  } + c.c.  \right] 
\end{eqnarray}
Note the appearance of an oscillating phase in the quadrature $ e^{i \Delta \phi} $, with 
\begin{eqnarray}
\Delta \phi = ( q \lambda) \sin \Omega_{GW} t    
\end{eqnarray}
proportional to the quantity $  q \lambda $.
We wish to express the value of $  q \lambda $ in terms of experimentally accessible quantities. The energy density of a gravitational wave is \cite{Landau1975}
\begin{eqnarray}
E = (1 / 32\pi G ) \Omega_{GW}^{2} f^{2}
\end{eqnarray}
where $ f $ is the wave's strain. The energy density of a single graviton of frequency $ \Omega_{GW} $ is
\begin{eqnarray}
E_{g} = \dfrac{\Omega_{GW}}{V}
\end{eqnarray}
For a coherent state we have that $ \lambda = \sqrt{ \langle N \rangle} $, where $  \langle N \rangle $ is the mean number of gravitons in the state. Moreover, the mean number of gravitons in a coherent gravitational wave can also be written as
\begin{eqnarray}
\sqrt{  \langle N \rangle } = \sqrt{\dfrac{E}{E_{g}}} = \sqrt{\dfrac{V \Omega_{GW}}{32\pi G}}  f
\end{eqnarray}
and combining the expressions,
\begin{eqnarray}
\Delta \phi = \left(  \dfrac{\omega_{0}}{\Omega_{GW}} \right)  f  \sin \Omega_{GW} t    
\end{eqnarray}
This is the same as obtained using classical general relativity. The phase of the optical field oscillates at the frequency of the gravitational wave, with an amplitude proportional to the wave's strain. For an optical frequency, a gravitational wave frequency of \SI{100}{Hz} and a strain of $ f \sim 10^{-21} $, the typical frequency range and strain of waves detected by LIGO, oscillations in the phase are on the order of one part in $ 10^{9} $. A classical back-of-the-envelope estimate yields $ \Delta \phi = b \times (2\pi \ell_{0} / \lambda) f \approx 10^{-9}$, where $ b \approx 200 $ is the number of bounces of the photon within the interferometer \cite{LIGO2019}. 

\section{Squeezed states}

One interesting calculation, which cannot be done within the classical theory, is the effect of a squeezed gravitational wave on the cavity field. Squeezed gravitational states, which are quantum mechanical configurations of the gravitational field, are important in cosmology. It is well known that the Big Bang leaves a Cosmic Gravitational Wave Background (GWB) and theories of inflation predict that relic gravitational waves can be squeezed \cite{Allen1996}. 

Consider once again an optical coherent state $ \vert \alpha \rangle $ and a squeezed gravitational wave state $ \vert \psi \rangle = \vert \xi_{0} e^{i \theta} \rangle $, where $ \alpha, \xi_{0} $ are chosen real numbers for simplicity and $ \theta = 2\Omega_{GW}t $ is the time dependency of the squeezed state. We can write the squeezed state as a squeezing operator acting on the vacuum state, $  \vert \xi_{0} e^{i \theta} \rangle = \mathcal{S} \vert 0 \rangle $. In the approximation that the ``Kerr-like'' quadratic term of the evolution operator and the vacuum fluctuations of the gravitational field are negligible, the mean electric field is
\begin{eqnarray}
\langle \mathcal{\tilde{E}_{+}} \rangle = \alpha \langle 0 \vert \mathcal{S}^{\dagger} \mathcal{D}(- q \gamma) \mathcal{S}  \vert 0 \rangle + c.c.
\end{eqnarray}
Assuming $ \xi_{0} \gg 1 $ (super-horizon scales \cite{japs}), the operator $ \mathcal{S}^{\dagger} \mathcal{D} \mathcal{S} $ is well approximated by a displacement operator $ \mathcal{D}(-q  e^{\xi_{0}} (\gamma + \gamma^{*} e^{i\theta}) ) $, from which the vacuum-vacuum amplitude can be readily calculated. The electric field strain becomes
\begin{eqnarray}
\langle \mathcal{E}_{+} \rangle &=& \alpha \exp \left(  -\dfrac{1}{2}  \vert  -q  e^{\xi_{0}} (\gamma + \gamma^{*} e^{i\theta})  \vert^{2}  \right) + c.c. \\
&\simeq & 2 \alpha \left(    1 -  8 q^{2}e^{2\xi_{0}} \sin^{4} \left(  \dfrac{\Omega_{GW}t}{2}  \right)    \right) 
\end{eqnarray}
Once again, we may write the above expression in terms of experimentally measurable quantities. The squeezing parameter can be related to the gravitational wave frequency  \cite{Allen1996, japs} by $ e^{\xi_{0}}  \approx \sinh \xi_{0} = ( \upsilon / \Omega_{GW} )^{2} / 2 $, where 
\begin{equation}
\upsilon = 10^{9} \sqrt{\dfrac{H}{10^{-4} M_{\rm pl}}} \  (\SI{}{Hz})
\end{equation}
and thus
\begin{eqnarray}
\langle \mathcal{E}_{+} \rangle \simeq 2 \alpha \left(  1   -8 \left(   \dfrac{\omega_{0}}{E_{\rm pl}} \right)^{2}  \left(    \dfrac{\upsilon}{\Omega_{GW}} \right)^{4}  \sin^{4} \left(  \dfrac{\Omega_{GW} t}{2}  \right) \right) \nonumber
\\
\ \label{oscillations_squeeze}
\end{eqnarray}
For $ H = 10^{-4} M_{\rm pl} $ and $ \Omega_{GW} = \SI{0.1}{Hz} $ the pre-factor in the oscillating part of  \eqref{oscillations_squeeze} is approximately $ 10^{-15} $, which could be probed within future experiments.
Measuring the oscillation in the quadrature predicted by equation \eqref{oscillations_squeeze} could provide evidence on the quantum mechanical nature of gravity and help establish cosmological models predicting the occurrence of squeezed gravitational waves.

\section{Conclusion}

We have discussed the quantum mechanical interaction of gravitational waves with an optical cavity. For the gravitational field in the vacuum state, the cavity field experiences squeezing and gravitationally induced revivals of squeezing. Measuring these revivals would provide evidence on the quantum nature of the gravitational field. Estimates show, however, that gravitationally induced squeezing is far beyond the reach of low energy experiments. 

We have also considered the quantum treatment of a coherent gravitational wave interacting with the cavity. We were then able to retrieve the result calculable within classical general relativity, but using quantum mechanical methods.  

One calculation we can perform, which cannot be realized within the classical theory is the interaction of a squeezed gravitational wave with the cavity field. We have shown that in that case the field quadrature experiences oscillations at half the frequency of the wave. This could be important in the study of cosmological models predicting squeezed gravitational waves, and could provide indirect evidence on the quantum nature of the gravitational field. 

In summary, by monitoring the electric field inside a cavity it is in principle possible to probe quantum effects of the gravitational field. 
Quantum gravity has been deemed outside the realm of experiments. One must not forget, however, that physics is an experimental science and that it is quite possible that the quantum nature of gravity can be probed with alternative quantum experiments after enough effort has been deployed.

\begin{acknowledgments}
\textit{Acknowledgements.} The author acknowledges Bruno Melo, George Svetlichny and Carlos Tomei for stimulating discussions and the anonymous referee whose comments greatly improved the paper. This study was financed in part by the Coordena\c{c}\~ao de Aperfei\c{c}oamento de Pessoal de N\'ivel Superior - Brasil (CAPES) - Finance Code 001.
\end{acknowledgments}


\onecolumngrid

\appendix

\section*{Appendix: Derivation of the multi-mode unitary operator}

In this appendix we provide a derivation of equations \eqref{unitary} and \eqref{Uk} used in the main text. This is similar to the derivation provided in \cite{Bose1997}, generalized here to the multi-mode case. 
Start by considering the Hamiltonian 
\begin{eqnarray}
\hat{H} = \omega_{0} a^{\dagger}a +\sum_{k}  \Omega_{k} b^{\dagger}_{k} b_{k} - \sum_{k} g_{k} a^{\dagger} a \ (b_{k}  + b^{\dagger}_{k})
\end{eqnarray}
The evolution operator reads
\begin{eqnarray}
U(t) = e^{-i\hat{H}t }
\end{eqnarray}
Note that since $ \left[   b_{k},   b_{k'}  \right]  = 0  \ , \  k\neq k' $ we may write 
\begin{eqnarray}
U(t) = e^{-i\omega_{0} a^{\dagger}a} \prod_{k} U_{k} (t) 
\end{eqnarray}
in which
\begin{eqnarray}
U_{k} (t)  &=& \exp \left\lbrace   -it\Omega_{k} b^{\dagger}_{k} b_{k} +it  g_{k} a^{\dagger} a \ (b_{k}  + b^{\dagger}_{k})    \right\rbrace \\
&=& \exp \left\lbrace   -i \tilde{t} b^{\dagger}_{k} b_{k} +i \tilde{t}  q_{k} a^{\dagger} a \ (b_{k}  + b^{\dagger}_{k})    \right\rbrace 
\end{eqnarray}
where we have re-scaled the time variable $ \tilde{t} = \Omega_{k}t $. To obtain the desired expression, we must make repeated use of the well-known Baker-Campbell-Hausdorff (BCH) formula and of the fact that for an operator $ A $ and a function $ f(A) $,
\begin{eqnarray}
V f(A) V^{\dagger} = f(\left\lbrace V A V^{\dagger}  \right\rbrace )
\end{eqnarray}
Define for each mode $ k $ the unitary operator
\begin{eqnarray}
V_{k} = e^{-q_{k} a^{\dagger}a (b_{k} - b^{\dagger}_{k})}
\end{eqnarray}
Using BCH we can show that:
\begin{eqnarray}
V_{k} b_{k} V_{k}^{\dagger} &=&  b_{k} + q_{k} a^{\dagger}a
\\
V_{k} b^{\dagger}_{k} b_{k} V_{k}^{\dagger} &=& b^{\dagger}_{k} b_{k} + q_{k} a^{\dagger}a ( b^{\dagger}_{k} +  b_{k} ) + q_{k}^{2} (a^{\dagger}a )^{2}
\\
V_{k} a^{\dagger} a V_{k}^{\dagger} &=&  a^{\dagger} a
\end{eqnarray}
It is simple to show that 
\begin{eqnarray}
V_{k}  U_{k} V_{k}^{\dagger}  = e^{-i\tilde{t}(b^{\dagger}_{k} b_{k})} e^{i q_{k}^{2}(a^{\dagger} a )^{2}}
\end{eqnarray}
To place $ U_{k} $ in the desired form, we simply multiply the above expression by $ V_{k}^{\dagger} $ on the left and $ V_{k}  $ on the right. By direct calculation,
\begin{eqnarray}
U_{k} (t)  = e^{iA_{k}(t)(a^{\dagger}a)^{2}} e^{q_{k} a^{\dagger} a  ( \gamma_{k}^{*} b_{k} - \gamma_{k} b^{\dagger}_{k}   )}
\end{eqnarray}
where $ \gamma_{k} = (1 - e^{i\Omega_{k} t}) $ and $ A_{k}(t) = 2 q_{k}^{2} \left(   \Omega_{k} t - \sin \Omega_{k} t  \right)   $. This is the form used in the main text.

\end{document}